\documentclass[a4paper, oneside, twocolumn, notitlepage, 10pt]{extarticle_ecoc}
\usepackage{ecoc}

\usepackage[hidelinks]{hyperref}
\providecommand*{\thesection}{1}

\usepackage{cleveref}
\usepackage{tikz}
\usepackage{orcidlink}
\usetikzlibrary{positioning, calc, shapes.geometric}
\usepackage[dvipsnames]{xcolor}
\usepackage{pgfplots}
\usepackage{pgfplotstable} 
\pgfplotsset{compat=1.18}
\usepackage[per-mode = symbol]{siunitx}
\usepackage{subcaption}
\DeclareSIUnit\sample{Sa}
\pgfplotsset{compat=1.18}
\usetikzlibrary{arrows.meta, decorations.pathmorphing, patterns, shapes.geometric, calc}
\tikzset{dashed with dot/.style={line cap=round,line width=2pt,dash pattern=on 0pt off 2.5pt}}
\usepgfplotslibrary{groupplots}
\usepackage[dvipsnames,svgnames,x11names]{xcolor}
\usepgfplotslibrary{fillbetween}
\usepackage[justification=justified]{caption}
\usepgfplotslibrary{statistics}
\usepgfplotslibrary{fillbetween}
\usetikzlibrary{patterns}
\usetikzlibrary{patterns.meta}
\usetikzlibrary{backgrounds}
\usetikzlibrary{spy}

\crefname{figure}{Fig.}{Figs.}
\crefname{table}{Tab.}{Tabs.}

\begin{document}
\selectlanguage{english}    % Standard Language

%-------------------------------------------------- Title -----------------------------------------------------%

% \title{Accurate Localization of Multiple Simultaneous Perturbations in Optical Fibers via Polarization Sensing at the Receiver}
\title{Accurate Multi-perturbation Localization in Optical Fibers with Polarization-based Forward Sensing}

%Title: Please Capitalise the First Letter of Each Principal Word\\(Except "and", "of", "the", etc.)
%------------------------------------------------- Authors-----------------------------------------------------%

\author{
    Lampros Lanaras\textsuperscript{(1)},
    Rick M. Butler\textsuperscript{(2,1,*)},
    Christian H\"ager\textsuperscript{(2)},
    and Alex Alvarado\textsuperscript{(1)}
}

\maketitle                  % Create title and author

%------------------------------------------ Description of Authors ----------------------------------------------%

\begin{strip}
    \begin{author_descr}

        \textsuperscript{(1)} Department of Electrical Engineering, Eindhoven University of Technology, Eindhoven, the Netherlands

        % \textcolor{blue}{\uline{author@institution.org}} (Email address of corresponding author mandatory)

        \textsuperscript{(2)} Department of Electrical Engineering, Chalmers University of Technology, Gothenburg, Sweden
        % \textcolor{blue}{\uline{author@institution.org}} (Email address optional)

        % \textsuperscript{(3)} Authors' full affiliation,
        % \textcolor{blue}{\uline{author@institution.org}} (Email address optional)
                \textsuperscript{(*)}\textcolor{blue}{\uline{rick.butler@chalmers.se}}

    \end{author_descr}
\end{strip}

% \setstretch{1.1}
%-------------------------------------------------- Footnote -------------------------------------------------------%
\renewcommand\footnotemark{}
\renewcommand\footnoterule{}
%\let\thefootnote\relax\footnotetext{text}

%-------------------------------------------------- Abstract ---------------------------------------------------------%

\begin{strip}
    \begin{ecoc_abstract}
    %Problem statement
    % Current optical fiber sensing at the receiver can localize one perturbation at a time.
    We present a method to accurately localize multiple perturbations from fiber outputs.
    % Methods
    We cross-correlate received polarizations between two wavelengths, apply peak detection, and interpolate peaks to boost accuracy.
    % Results, discussion, conclusion
    We localize five simultaneous perturbations in a \qty{10}{\giga\sample\per\second} numerical fiber model with \qty{3.5}{\meter} median error.
    % Copyright
    ©2026 The Author(s)
    \end{ecoc_abstract}
\end{strip}

\section{Introduction}
% % Beyond communications, the optical fiber infrastructure can be leveraged for sensing applications, such as real-time seismic activity detection \cite{Awad:May24:environmental_surveillance_networks} and critical infrastructure surveillance \cite{Fiber_distributed_sensing}.
% Beyond communications, the optical fiber infrastructure can be leveraged for sensing \cite{distr_perturbations,Fiber_distributed_sensing,trens_secure_network}.
% % At the same time, the ever-expanding dependence on these networks for global communications \cite{un_fibers_are_critical} creates a critical need to monitor fiber health \cite{trens_secure_network}.
% Societal and economic relevance drive worldwide initiatives %e.g. European initiatives \cite{ec_submarine_cables_2026} 
% to promote the development of perturbation sensing technologies, e.g., \cite{ec_submarine_cables_2026}.

Distributed acoustic sensing (DAS) localizes fiber perturbations from backscatter phase at the transmitter \cite{DAS}, but it requires expensive equipment.
An alternative is to monitor the state of polarization (SOP) of propagated waves, often called forward polarization sensing \cite{Mecozzi}.
External perturbations change the fiber cross-section geometry, scrambling the SOP.
Although these SOP variations can be measured in a standard coherent receiver, they lack the inherent localization ability to DAS.

Distributed forward sensing schemes were proposed in \cite{Tang,Vaskin,Chen:26:intrinsic_distributed_sensing} that use chromatic dispersion (CD) to localize perturbations.
Two waves propagate at distinct wavelengths while the fiber is perturbed.
In \cite{Tang,Vaskin}, the receiver observes the two resulting SOP variations with a CD-induced delay, from which the perturbation can be localized directly.
\cite{Chen:26:intrinsic_distributed_sensing} monitors global phase changes instead.
The sampling rate determines how finely one can determine this delay, and hence the localization resolution.

The method in \cite{Tang,Vaskin} senses one perturbation at a time, using a single pair of SOP variations.
In practice, perturbations can occur simultaneously or distributed in space \cite{distr_perturbations,Fiber_distributed_sensing,trens_secure_network}.
Additionally, the localization resolution of the methods in \cite{Tang,Vaskin,Chen:26:intrinsic_distributed_sensing} is limited by the sample rate.
Indeed, the results in \cite{Tang} show sub-\SI{25}{\meter} mean error using a \SI{32}{\giga\sample\per\second} coherent receiver, \cite{Vaskin} reported sub-\SI{5.4}{\kilo\meter} mean errors with a
\SI{100}{\mega\sample\per\second} polarimeter, and \cite{Chen:26:intrinsic_distributed_sensing} demonstrated \qty{1.57}{\kilo\meter} error with a \qty{625}{\mega\sample\per\second} receiver.

In this paper, we extend the method from \cite{Tang,Vaskin} to sense multiple perturbations at high resolution, without requiring high sampling rates.
We cross-correlate received SOP angular speeds (SOPASs) at two wavelengths, and apply a multi-peak detector to detect an arbitrary number of perturbations.
We interpolate each peak to find delays at noninteger multiples of the sampling period, which greatly improves the localization resolution.
% Then, we follow the method from \cite{Tang, Vaskin} and translate the time occurrence of each peak to a location along the fiber using CD.
To the best of our knowledge, this is the first forward sensing method that localizes multiple perturbations using CD and SOP, and boosts accuracy using interpolation.
In numerical simulations of a linear \SI{50}{\kilo\meter} single-span fiber and two carriers within the C-band sampled at \SI{10}{\giga\sample\per\second}, we demonstrate localization of five simultaneous perturbations with a median error of \SI{3.5}{\meter}.
% To the best of our knowledge, this is the first demonstration of such accurate distributed forward sensing at the receiver using CD.
% Furthermore, we show that SOP angular speed (SOPAS) \cite{Pellegrini_SOPAS} and cross-correlation can be leveraged to localize multiple near-simulateneous perturbations in the fiber model using CD.
% We demonstrate localization of multiple simultaneous, spatially separated perturbations for the first time.
% We verify our method using synthetic data from a numerical fiber model model which incorporate perturbation effects on DGD, attenuation, and optical noise.

%excludes nonlinearities, pulse spreading, or polarization-dependent loss.

%%%%%%%%%%%%%%%%%%%%%%%%%%%%%%%%%%%%%%%%%%%%

\section{Perturbed propagation model}
\begin{figure*}[tb]
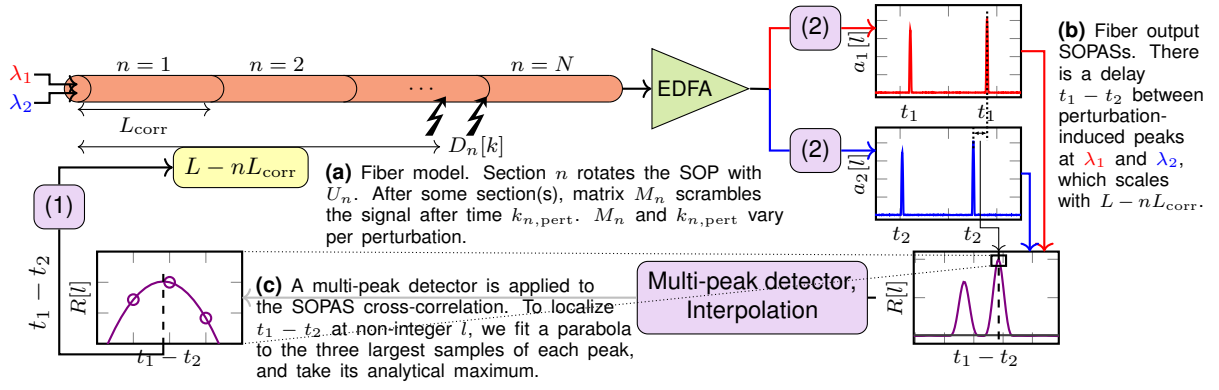

\centering
\newlength{\nodeDistance}
\newlength\fibreRadius
\newlength\sectionLength
\newlength\plotSeparation
\newlength\plotHeight
\newlength\plotWidth

\setlength{\nodeDistance}{.4cm}

% Model parameters
\edef\colourOne{red}
\edef\colourTwo{blue}
\setlength\fibreRadius{.5em}
\setlength\sectionLength{5em}
\edef\sectionCount{4}
\pgfmathtruncatemacro{\sectionCountMinusOne}{\sectionCount - 1}
\pgfmathtruncatemacro{\sectionCountPlusOne}{\sectionCount + 1}
\edef\fibreEdgeColour{Melon}

% SOPAS parameters
\setlength\plotSeparation{.4cm}
\setlength\plotHeight{2.8cm}
\setlength\plotWidth{3.5cm}

\begin{tikzpicture}[
        x = 1cm,
        y = 1cm,
        node distance = \nodeDistance,
        field/.style = {
            align = center,
            anchor = center,
            inner sep = 0
        },
        zoom/.style = {
            field,
            draw = black,
            line width = .7pt
        },
        caption/.style = {
            field,
            anchor = north,
        },
        block/.style = {
            field,
            draw = black,
            rounded corners,
            font = \small
        },
        arrow/.style = {
            ->,
            draw = black
        },
        fibre/.style = {
            line width = \fibreWidth
        },
        earthquake/.style = {
            draw = \earthquakeColour,
            fill = \earthquakeColour,
            line width = \earthquakeRingWidth
        }
    ]
    
    % Draw the subfigures
    \begin{scope}[local bounding box = Fibre, shift = {(0, 0)}]
        \input{Figures/fibre}
    \end{scope}
    
    \begin{scope}[local bounding box = Traces, shift = {(11.3, -.5)}]
        \input{Figures/SOPAS}
    \end{scope}

    \node[block, rounded corners, inner sep = 4pt, fill = DarkOrchid!20!white] at ($(SOPAS1.north west)!.5!(SOPAS1.west) + (-.8, 0)$) (SOPAS1Calculation) {\eqref{eq:sopas}};
    \node[block, rounded corners, inner sep = 4pt, fill = DarkOrchid!20!white] at ($(SOPAS2.north west)!.5!(SOPAS2.west) + (-.8, 0)$) (SOPAS2Calculation) {\eqref{eq:sopas}};

    \begin{scope}[local bounding box = Correlation, shift = {(11.8, -2.95)}]
        \input{Figures/crosscorrelation}
    \end{scope}

    \node[block, at = (CrossCorrelation.west), anchor = east, xshift = -1.725em, fill = DarkOrchid!20!white, inner sep = 4pt] (Interpolation) {Multi-peak detector,\\Interpolation};

    \begin{scope}[local bounding box = CorrelationInterpolated, shift = {(1, -2.95)}]
        \input{Figures/crosscorrelation_interpolated}
    \end{scope}

    \node[block, fill = DarkOrchid!20!white, inner sep = 4pt] at (.5, -1.75) (Estimator) {\eqref{localization_equation}};

    % Draw connections
    \coordinate (SOPASMidpoint) at ($(SOPAS1Calculation.west)!.5!(SOPAS2Calculation.west)$);
    \coordinate (SOPASPath) at ($(EDFA.east)!.5!(SOPASMidpoint)$);
    \draw[thick] (EDFA.east) -- (SOPASPath);
    \draw[thick, \colourOne] (SOPASPath) |- (SOPAS1Calculation.west);
    \draw[thick, \colourTwo] (SOPASPath) |- (SOPAS2Calculation.west);
    \draw[arrow, thick, \colourOne] (SOPAS1Calculation) -- (SOPAS1Calculation.east -| SOPAS1.west);
    \draw[arrow, thick, \colourTwo] (SOPAS2Calculation) -- (SOPAS2Calculation.east -| SOPAS2.west);

    \coordinate (CorrelationInput1) at ($(CorrelationNorthWest)!.1!(CorrelationSouthWest)$);
    \coordinate (CorrelationInput2) at ($(CorrelationNorthWest)!.9!(CorrelationSouthWest)$);
    \coordinate (CorrelationMidpoint1) at ($(SOPAS1.east)!.5!(CorrelationInput1)$);
    \coordinate (CorrelationMidpoint2) at ($(SOPAS2.east)!.5!(CorrelationInput2)$);
    \draw[arrow, thick, \colourOne] (SOPAS1.east) -| ($(CrossCorrelation.north west)!.9!(CrossCorrelation.north east)$);
    \draw[arrow, thick, \colourTwo] (SOPAS2.east) -| ($(CrossCorrelation.north west)!.8!(CrossCorrelation.north east)$);

    \draw[arrow] (SOPASPeakDifference.south) -- ++(0, -1.15) -| (CorrelationPeakAnchor |- CrossCorrelation.north);

    \draw[densely dotted] (ROINorthWest) -- (CrossCorrelationInterpolated.north east);
    \draw[densely dotted] (ROISouthWest) -- (CrossCorrelationInterpolated.south east);
    \draw[thick] ($(CrossCorrelation.west) + (-.5, 0)$) -- (Interpolation);
    \draw[arrow, thick, lightgray] (Interpolation) -- (CrossCorrelationInterpolated.east);

    \draw[thick] (CorrelationInterpolatedPeak |- CrossCorrelationInterpolated.south) -- ++(0, -.4em) -| (Estimator) node[pos = .75, sloped, anchor = south] {$t_1 - t_2$};
    \draw[arrow, thick] (Estimator) |- (PerturbationDistance);

    % Subcaptions
    \node[caption, at = (Perturbation1), anchor = north, text width = .38\textwidth, xshift = 3em] {
        \subcaption{
            Fiber model.
            Section $n$ rotates the SOP with $U_n$.
            After some section(s), matrix $M_n$ scrambles the signal after time $k_{n,\textrm{pert}}$.
            $M_n$ and $k_{n,\textrm{pert}}$ vary per perturbation.
        } \label{fig:schematic:fiber}
    };

    \node[caption, at = (SOPAS1.north east), anchor = north west, xshift = 1.4em, text width = .12\textwidth] {
        \subcaption{
            Fiber output SOPASs.
            There is a delay $t_1 - t_2$ between perturbation-induced peaks at \textcolor{\colourOne}{$\lambda_1$} and \textcolor{\colourTwo}{$\lambda_2$}, which scales with $L - nL_\textrm{corr}$. 
        } \label{fig:schematic:SOPAS}
    };

    \node[caption, at = (Interpolation.south), anchor = north, xshift = -11.6em, yshift = 2.8em, text width = .31\textwidth] {
        \subcaption{
            A multi-peak detector is applied to the SOPAS cross-correlation.
            To localize $t_1 - t_2$ at non-integer $l$, we fit a parabola to the three largest samples of each peak, and take its analytical maximum.
        } \label{fig:schematic:detector}
    };
\end{tikzpicture}
\vspace{1ex}
\caption{
    System block diagram.
    (a) While the carriers propagate, the fiber model is perturbed twice.
    (b) We correlate the received SOPASs $a_1[l]$ and $a_2[l]$.
    (c) After multi-peak detection and interpolation, we estimate $L - nL_\textrm{corr}$ for both perturbations.
} \label{fig:schematic}
\end{figure*}

We model two propagating dual-polarization carrier waves using the coarse-step method \cite{Curti:Aug90:waveplate}, as visualized in \cref{fig:schematic:fiber}.
A length-$L$ fiber is divided into $N = L / L_\textrm{corr}$ sections, where $L_\textrm{corr}$ is the correlation length.
The noncircular fiber cross-section causes two orthogonal polarizations to propagate faster- and slower than all others in section $n$.
To randomize these polarizations and their speed difference, section $n$ scrambles the SOP via a random special unitary matrix $U_n \in \mathbb{C}^{2\times 2}$.
% $U_n$ randomizes the DPS, and rotates the fastest and slowest polarizations.
A random set $\left\{U_n : n = 1, \dots, N\right\}$ defines a fiber realization.
We generate many realizations to model SOP scrambling statistics over many fibers.
Following \cite{Czegledi_SOP}, we generate $U_n$ as a random-uniform SOP rotation, assuming that any fiber cross-section geometry is equally likely.
Additionally we model CD, attenuation and amplifier noise while excluding the Kerr nonlinearity and polarization-dependent loss.

To model a sudden perturbation at the end of section $n$, we replace $U_n$ by a matrix $D_n[k]U_n$ where $D_n[k] \in \mathbb{C}^{2\times 2}$ is the perturbation matrix at time $k\Delta t_\textrm{pert}$, and $\Delta t_\textrm{pert}$ is the sample period.
Given that the perturbation occurs at sample $k_{n,\textrm{pert}}$, $D_n[k]$ is the identity matrix if $k < k_{n,\textrm{pert}}$, and unitary matrix $M_n \in \mathbb{C}^{2\times 2}$ otherwise.
Like $U_n$, we draw $M_n$ random-uniformly, assuming that any perturbation-induced scrambling is equally likely.

We model propagating waves with a sample period $\Delta t_\textrm{wave} \neq \Delta t_\textrm{pert}$.
Therefore, the time at which a propagating sample reaches the end of section $n$, can occur between sample times of $D_n[k]$. %at a noninteger $k$.
In these cases, we interpolate the propagation matrices at the two integer values closest to $k$, using piecewise cubic interpolation \cite{PCH_interpolation}.
This simulates a fast but noninstantaneous perturbation.
Additionally, to emphasise the different sample rates, we denote perturbation time samples with $k$ and wave time samples with $l$.
% Sample $l$ is perturbed by matrix $D(t_n[l])$.
% If the corresponding perturbation sample $k_n[l] = t_n[l] / \Delta t_\textrm{dist}$ is no integer, we interpolate $D_n(\lceil k_n[l]\rceil\Delta t_\textrm{dist})$ and $D_n(\lfloor k_n[l]\rfloor\Delta t_\textrm{dist})$ to $D_n(k_n[l]\Delta t_\textrm{dist})$ using piecewise cubic interpolation \cite{PCH_interpolation}.
% When $-1 < k_n[l] - k_\textrm{dist} < 0$, this simulates a fast but noninstantaneous scrambler.
% % These perturbations manifest as near-instantaneous steps in the SOP, and are evident in the evolution of the output Stokes parameters.

\section{Localization method}
\Cref{fig:schematic:SOPAS} shows the CD-based localization principle that \cite{Tang,Vaskin} are based on.
% When the two signals propagate through the fiber during a perturbation, their SOPAS traces contain the same spike at different temporal positions.
We consider two waves with wavelengths $\lambda_1$ and $\lambda_2$.
% When the fiber is perturbed, this causes simultaneous spikes in their SOPASs $a_1[k]$ and $a_2[k]$ at that location.
A sudden perturbation $D_n[k]$ changes their SOP after section $n$.
Given group velocities $v_{g_1} < v_{g_2}$, these SOP changes reach the receiver at times $t_1 = (L - nL_\textrm{corr}) / v_{g_1}$ and $t_2 = (L - nL_\textrm{corr}) / v_{g_2}$ after the event.
% We denote this delay as $\Delta t_{rel}$.
The further the waves propagate after the perturbation, the further apart $t_1$ and $t_2$ will be.
% If $a_1[l]$ and $a_2[l]$ are the SOPAS traces of two signals, we identify the sample delay $\tau$ corresponding to a perturbation, by locating the respective cross-correlation peak.
% The relative time delay $\Delta t_{rel}$ accumulated from the perturbation location up to the receiving end is then computed as  $\Delta t_{rel}=\tau / R_s$.
Specifically, the distance of the perturbation to the receiver is \cite{Tang}
\begin{equation} \label{localization_equation}
    L - nL_\textrm{corr} = \frac{(t_1 - t_2)v_{g_1} v _{g_2}}{v_{g_2}-v_{g_1}}.
\end{equation}
% We localize the perturbation with \cite{Tang}

% Localization is achieved through the wavelength dependence of the signal group velocities.
Before evaluating \eqref{localization_equation}, we must obtain $t_1 - t_2$ at the receiver.
Expressing the fiber output SOP at time $l\Delta t_\textrm{wave}$ as Stokes vector $\vec{S}_1[l] = \begin{bmatrix}S_{1, 1}[l], S_{1, 2}[l], S_{1, 3}[l]\end{bmatrix}^T$, we measure SOP fluctuations via SOPAS \cite{Pellegrini_SOPAS}
\begin{equation} \label{eq:sopas}
  a_m[l] = \frac{1}{\Delta t_\textrm{wave}}\left|\arccos\Bigg (\frac{ \vec{S}_m[l] \cdot\vec{S}_m[l-1]}{\| \vec{S}_m[l]\| \|\vec{S}_m[l-1]\|}\Bigg)\right|.
\end{equation}
% SOPAS reduces the three-dimensional SOP fluctuations to a one-dimensional metric for simplicity.
% With SOPAS, we are able to reduce the problem dimensionality and transform near-instantaneous rotations into narrow spikes.
As the derivative of $D_n[k]$ is nonzero for a $\Delta t_\textrm{pert}$-period, it causes a $\Delta t_\textrm{pert}$-wide peak in $a_m[l]$.
Now, cross-correlation $R[l] = a_1[l] \star a_2[l]$ has a $2\Delta t_\textrm{pert}$-wide peak centered at $t_1 - t_2$ \cite{Tang}, as demonstrated in \cref{fig:schematic:SOPAS}.

\section{Accurately localizing multiple perturbations}
Maximizing $R[l]$ yields one peak, and hence, localizes one perturbation.
To localize multiple perturbations, we propose to detect all local maxima in $R[l]$ above a $4b\hat\sigma$ threshold as visualised in \cref{fig:schematic:detector} (right), where $b$ is a constant, $\hat\sigma = \tilde\mu_l(|R[l] - \tilde\mu_m(R[m])|)$, and $\tilde\mu_x(R[x])$ calculates the median of $R[x]$ over all integers $x$ \cite{MAD}.

$M$ perturbations would cause $M$ peaks in both $a_1[l]$ and $a_2[l]$.
Consequently, $R[l]$ would contain up to $M^2$ peaks.
Of these, only up to $M$ can be true positives (TPs).
To filter out false positives (FPs), we define $\Delta t_{\textrm{max}}$ as the largest possible $t_1 - t_2$ that can be a TP.
$\Delta t_{\textrm{max}}$ corresponds to a perturbation at the transmitter.
Thus, we discard all peaks at $l\Delta t_\textrm{wave} > \frac{L}{v_{g_1}} - \frac{L}{v_{g_2}}$.
Now, for a FP to still occur, two perturbations must be timed and/or spaced sufficiently close that the resulting peak in $R[l]$ satisfies this constraint.
% Luckily, if perturbations are far apart in space, the $M^2 - M$ false positive peaks occur at relatively large values of $l$.
% These peaks can be discarded by thresholding $l$.
% However, to make the algorithm robust to spatially close perturbations, another solution is necessary.

Discrete peak detection can detect peaks in $R[l]$ only at multiples of $\Delta t_\textrm{wave}$.
This imposes a strong dependence of localization error on the sampling rate.
Assuming uniformly distributed quantization errors, quantization causes an expected error in the detected peak time of
\begin{equation}
    \frac{1}{\Delta t_\textrm{wave}}\int_{-\Delta t_\textrm{wave} / 2}^{\Delta t_\textrm{wave} / 2} |x| \mathrm{d}x
    % = 
    % \frac{2}{\Delta t_\textrm{wave}}\int_0^{\Delta t_\textrm{wave} / 2} x \partial x
    % =
    % \frac{2}{\Delta t_\textrm{wave}}[0.5x^2]_0^{\Delta t_\textrm{wave} / 2}
    % =
    % \frac{2}{\Delta t_\textrm{wave}}(\Delta t_\textrm{wave}^2 / 8)
    =
    \frac{\Delta t_\textrm{wave}}{4},
\end{equation}
yielding an expected localization error in \eqref{localization_equation} of
\begin{equation} \label{eq:expected_error}
    \frac{\Delta t_\textrm{wave}}{4}\frac{v_{g_1}v_{g_2}}{v_{g_2}-v_{g_1}}.
\end{equation}
% as plotted in \cref{fig:results:error_vs_spacing}.
As this error is an artifact of discrete peak detection, we move to continuous peak detection at noninteger $l$.
Our approach is visualized in \cref{fig:schematic:detector}.
We parametrize the three samples of $R[l]$ closest to each peak as a parabola \cite{Parabolic_interpolation}, and place $t_1 - t_2$ at its analytical maximum.
% Because these samples are equally spaced, parabolic interpolation yields a closed-form algebraic expression, significantly reducing computational complexity compared to iterative fitting or resampling methods.

% The discrete nature of detecting the cross-correlation peak introduces a spatial uncertainty of $\pm 0.5$ samples.
% This uncertainty yields a quantization error which follows a uniform distribution over the interval $[-0.5/R_s, +0.5/R_s]$.
% Therefore, the absolute error is uniformly distributed over  $[0, +0.5/R_s]$, with a median value of $0.25/R_s$.
% calculated as $E[|e|] = \int_{-0.5/R_s}^{+0.5/R_s} |e| \cdot R_s \, de$, yields the theoretical resolution bound of $0.25/R_s$. 
% Substituting this value into \cref{localization_equation} yields a theoretical lower error bound of the median localization error:
% \begin{equation}
%     \mathrm{Median}(\textrm{error}) \geq L-\frac{v_{g_2}v_{g_1}}{4R_s(v_{g_2}-v_{g_1})}.
%     \label{eq:bound_equation}
% \end{equation}

% A dynamic threshold is applied to discard and record insufficient peaks.
% Specifically, we use a $4\hat{\sigma}$ threshold, based on the standard deviation estimator $\hat{\sigma}=b\cdot \mathrm{MAD}$. 
% Assuming normality of the data, the constant $b$ equals $1.4826$, while the median absolute deviation (MAD) of a particular trace $X$ is given by $\mathrm{MAD}= \mathrm{Median}(| x_i - \mathrm{Median}(X)|)$ \cite{MAD}. 

%\section{Wavelength spacing numerical experiment}
\section{Numerical results}

%This section and the next evaluate two numerical experiments on 
We consider a \SI{50}{\kilo\meter} fiber model with $N = \num{500}$, $L_\textrm{corr} = \SI{100}{\meter}$, attenuation \SI{0.2}{\decibel\per\kilo\meter}, and polarization mode dispersion parameter \SI{0.2}{\pico\second\per\sqrt{\kilo\meter}}.
An erbium-doped fiber amplifier (EDFA) (see \cref{fig:schematic:fiber}) is placed before the receiver.
%\SI{3}{\decibel} or \SI{6}{\decibel} in the first experiment, and \SI{6}{\decibel} in the second.
We transmit two $\num{50000}$-sample waves with $1 / \Delta t_\textrm{wave} = \SI{10}{\giga\sample\per\second}$, fixed $\lambda_2 = \qty{1565}{\nano\meter}$ and varying $\lambda_1$.
Perturbations are timed such that they are seen by both waves.
We set threshold constant $b = 1.4826$, taken from \cite{MAD}.
Each experiment is carried out over \num{50} fiber realizations.

\begin{figure}[tb]
    \centering
    \input{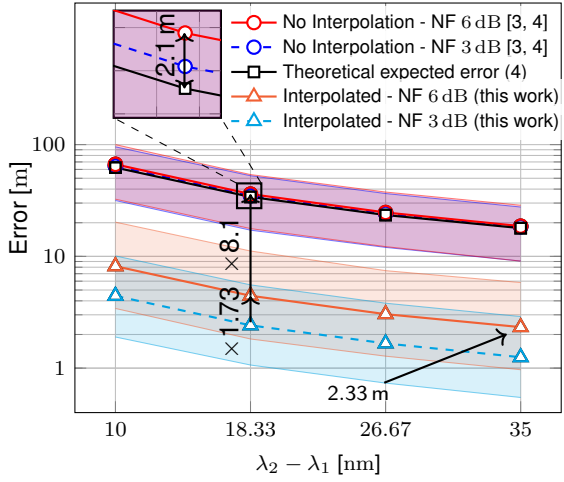}
    \caption{
    Median and interquartile range of the localization error for a single perturbation, as a function of $\lambda_2 - \lambda_1$ and EDFA noise.
    Results with and without our interpolation method are shown, as well as the expected localization error from \eqref{eq:expected_error}.
    % We show the cases amplifier noise figures of \SI{3}{\decibel} and \SI{6}{\decibel} are shown.
    % Each data point and its corresponding shaded area represent the median localization error and interquartile range respectively.
    }
    \label{fig:results:error_vs_spacing}
\end{figure}
%In the first numerical experiment, w
We first randomly perturb one section and 
% At each section, we evaluate $50$ independent fiber realizations with random birefringence by propagating a reference signal with a fixed carrier wavelength of \SI{1565}{\nano\meter}, and a second signal with a varied, shorter wavelength.
 localize this perturbation at the receiver using both the method from \cite{Tang,Vaskin} and our own.
We do this for amplifier noise figures (NFs) of $\SI{6}{\decibel}$ and $\SI{3}{\decibel}$, and %
% The median localization error is computed with and without interpolation of the cross-correlation peak, with amplifier noise figures at $\SI{3}{\decibel}$ and $\SI{6}{\decibel}$.
repeat this experiment for each possible perturbation location over a range of $\lambda_1$.
We report the localization error %median and inter-quartile range (IQR) 
as a function of $\lambda_2 - \lambda_1$.
% The results are compared against the theoretical lower error bound defined in \cref{eq:bound_equation}.

\Cref{fig:results:error_vs_spacing} shows this localization error median and interquartile range (IQR) as a function of $\lambda_2 - \lambda_1$.
As \cite{Tang} already observed and as expected from \eqref{eq:expected_error}, localization error decreases as $\lambda_2 - \lambda_1$ increases.
The error of \cite{Tang,Vaskin} approaches the expected localization error of \eqref{eq:expected_error}, although amplifier noise causes a mean offset of $\qty{2.1}{\meter}$.
Our interpolating method reduces the median error by at least a factor $8.1$, to \qty{2.33}{\meter} at $\lambda_2 - \lambda_1 = \qty{35}{\nano\meter}$.
For \qty{6}{\decibel} NF, the error increases by a factor \num{1.73} and by a factor of \num{1.04} for our method and for \cite{Tang,Vaskin}, respectively.
% The percentage of traces discarded by the threshold was less than $2\%$.
% A larger $\lambda_2 - \lambda_1$ increases the relative delay between the propagated signals and consequently, a higher sensing resolution is achieved, based on \cref{eq:expected_error}.
% With interpolation, \cref{eq:expected_error} is not applicable as the peak detection leverage sub-sample precision in computing $\tau$.

\Cref{fig:results:error_vs_spacing} shows that localization error in the original method from \cite{Tang,Vaskin} is dominated by quantization as described by \eqref{eq:expected_error}.
Although our method beats \eqref{eq:expected_error}, it still depends similarly on $\lambda_2 - \lambda_1$.
% We see from \eqref{localization_equation} that the effect of $t_1 - t_2$ on the estimate grows as $\lambda_2 - \lambda_1$ decreases.
This agrees with \eqref{localization_equation}, which shows that small mistakes in our peak fitting have larger consequences for smaller $\lambda_2 - \lambda_1$. Additionally, \Cref{fig:results:error_vs_spacing} shows that noise affects our method more than \cite{Tang,Vaskin}.
Any noise directly affects our parabolic peak fitting, whereas the quantized approach requires peak shifts of at least a half sample to increase localization errors.

%\section{Perturbation count numerical experiment}
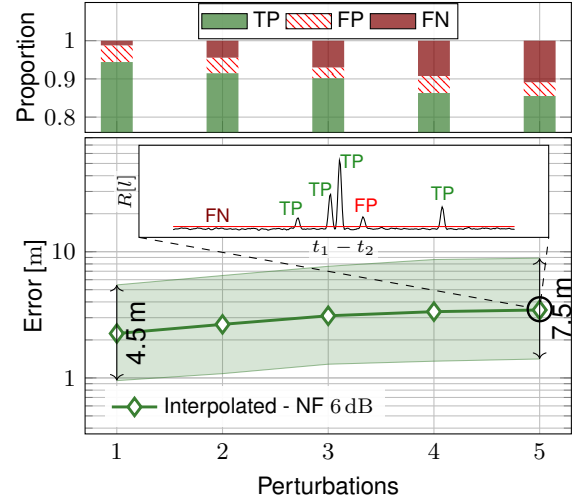
\begin{figure}[tb]
    \centering
    \begin{tikzpicture}
\pgfplotsset{set layers}
\pgfplotsset{all layers/.style={layer set=standard}}

\begin{groupplot}[
    group style={
        group size=1 by 2,
        group name=myplots,
        xlabels at=edge bottom,
        xticklabels at=edge bottom,
        vertical sep=2pt
    },
    width=1.05\columnwidth,
    grid=major,
    xlabel={Perturbations},
    xmin=0.7, xmax=5.3,
    xtick={1,2,3,4,5},
    xtick pos=bottom,
    ylabel style={yshift=-5pt, font=\small},
    xlabel style={font=\small},
    tick label style={font=\small}
]

% ==========================================
% TOP PLOT: Stacked Proportions (TP, FP, FN)
% ==========================================
\nextgroupplot[
    ybar stacked,
    height=3.3cm,
    bar width=12pt,
    ylabel={Proportion},
    ymin=0.76, ymax=1.1,
    ytick={0.8,0.9,1},
    area legend,
    legend columns=-1,
    legend style={
        at={(0.5,0.995)},
        anchor=north,
        font=\footnotesize,
        fill=white,
        fill opacity=0.9,
        text opacity=1,
        inner sep=1pt
    }
]

% true positives
\addplot[
    fill=OliveGreen,
    draw=none,
    fill opacity=0.7
] table[
    col sep=comma,
    x=actual_perts,
    y expr=\thisrow{total_true_matches}/(\thisrow{total_true_matches}+\thisrow{total_false_alarms}+\thisrow{total_missed_detections})
] {Figures/processed_monte_carlo_nlcap_global_median.csv};
\addlegendentry{TP}

% false positives
\addplot[
    draw=none,
    pattern=north west lines,
    pattern color=red,
    fill opacity=1
] table[
    col sep=comma,
    x=actual_perts,
    y expr=\thisrow{total_false_alarms}/(\thisrow{total_true_matches}+\thisrow{total_false_alarms}+\thisrow{total_missed_detections})
] {Figures/processed_monte_carlo_nlcap_global_median.csv};
\addlegendentry{FP}

% false negatives
\addplot[
    fill=Maroon,
    draw=none,
    % pattern=crosshatch,
    % pattern color=Maroon,
    fill opacity=0.7
] table[
    col sep=comma,
    x=actual_perts,
    y expr=\thisrow{total_missed_detections}/(\thisrow{total_true_matches}+\thisrow{total_false_alarms}+\thisrow{total_missed_detections})
] {Figures/processed_monte_carlo_nlcap_global_median.csv};
\addlegendentry{FN}

\nextgroupplot[
    ymode=log, 
    ylabel={Error [\unit{\meter}]},
    height=5.5cm,
    ymin=0.36,  
    ymax=80, 
    log ticks with fixed point,
    legend style={
        at={(0.04,0.161)},
        anchor=north west,
        font=\footnotesize,
        fill=white,
        fill opacity=0.93,
        text opacity=1,
        draw=none,
        inner sep=2pt,
    },
    grid=both,
    xtick style={draw=none},
    ylabel style={yshift=0pt, font=\small},
    xlabel style={font=\small},
    tick label style={font=\small}    
]

% Q1 line
\addplot[
    color=OliveGreen,
    opacity=0.5,
    name path=q1,
     forget plot
] table[
    col sep=comma,
    x=actual_perts,
    y expr=\thisrow{err_q1}
] {Figures/processed_monte_carlo_nlcap_global_median.csv};

% Q3 line
\addplot[
    color=OliveGreen,
    opacity=0.5,
    name path=q3,
     forget plot
] table[
    col sep=comma,
    x=actual_perts,
    y expr=\thisrow{err_q3}
] {Figures/processed_monte_carlo_nlcap_global_median.csv};

% Fill between Q1 and Q3
\addplot[
    OliveGreen,
    fill opacity=0.2,
    draw=none,
    forget plot
] fill between[of=q1 and q3];

% Median line
\addplot[
    color=OliveGreen,
    mark=diamond*,
    mark options={fill=white, solid, fill opacity=1},
    thick,
    line width=1.1pt,
    mark size=3.3pt
] table[
    col sep=comma,
    x=actual_perts,
    y expr=\thisrow{err_p50}
] {Figures/processed_monte_carlo_nlcap_global_median.csv};

\coordinate (q11) at (1, 0.9514);
\coordinate (q31) at (1, 5.4552);
\coordinate (q15) at (5, 1.4124);
\coordinate (q35) at (5, 8.8860);

\addlegendentry{Interpolated - NF \qty{6}{\decibel}}

% Define zoom point INSIDE the axis so axis cs works
\coordinate (zoompoint) at (axis cs:5,3.5);
\draw[black, line width=0.9pt] (zoompoint) circle[radius=4.5pt];

\end{groupplot}
%--------------------------------------------------------------------

% ==========================================
% COOL ARROWS
% ==========================================
\draw[<->] (q11) -- (q31) node[midway, sloped, anchor = north] {\normalfont \qty[detect-all]{4.5}{\meter}};
\draw[<->] (q15) -- (q35) node[midway, sloped, anchor = north, yshift = -.15em] {\normalfont \qty[detect-all]{7.5}{\meter}};

% ==========================================
% INSET PLOT
% ==========================================

\begin{axis}[
    name=insetplot,
    set layers=false,
    at={(myplots c1r2.south east)},
    anchor=south east,
    xshift=-0.3cm,
    yshift=2.6cm,
    width=7cm,
    height=2.8cm,
    xtick=\empty,
    ytick=\empty,
    ymin=-0.1,
    ymax=1.2,
    xlabel={$t_1 - t_2$},
    ylabel={$R[l]$},
    label style={
        font=\scriptsize,
        fill=white,
        inner sep=1pt,
        fill opacity=0.7,
        text opacity=1
    },
    title style={font=\footnotesize, yshift=-4pt},
    grid=none,
    axis background/.style={fill=white},
    clip=true
]
    \addplot[color=black,no marks]
        table[x=x, y=y, col sep=comma] {Figures/trace_seed10_5pert_realization_43.csv};

    \addplot[color=red]
        table[x=x, y=threshold, col sep=comma] {Figures/trace_seed10_5pert_realization_43.csv};

    \coordinate (insetSW) at (rel axis cs:0,0);
    \coordinate (insetSE) at (rel axis cs:1,0);

%%% peak labels
\node[anchor=south, font=\scriptsize, text=Maroon] at (axis cs: 90, -0.02) {FN};
\node[anchor=south, font=\scriptsize, text=ForestGreen] at (axis cs: 240, 0.08) {TP};
\node[anchor=south, font=\scriptsize, text=ForestGreen] at (axis cs: 300, 0.35) {TP};
\node[anchor=south, font=\scriptsize, text=ForestGreen] at (axis cs: 365, 0.75) {TP};
\node[anchor=south west, font=\scriptsize, text=Red] at (axis cs: 350, 0.13) {FP};
\node[anchor=south, font=\scriptsize, text=ForestGreen] at (axis cs: 550, 0.30) {TP};

\end{axis}
% ==========================================
% CONNECTOR LINES
% ==========================================
\draw[black, dashed, thin]
    ($(zoompoint)!0.04!(insetSW)$) -- (insetSW);

\draw[black, dashed, thin]
    ($(zoompoint)!0.10!(insetSE)$) -- (insetSE);

\end{tikzpicture}
    % \vspace{-4mm}
    \caption{
    % Median and interquartile range of the localization error for multiple perturbations.
    Localization metrics for multiple perturbations.
    (Top) The proportions of true positives, false positives, and false negatives.
    (Middle) Qualitative thresholded cross-correlation signal where five perturbations occurred.
    (Bottom) Median and interquartile range of the localization error.
    % The data points represent the median absolute localization error of valid matches, and the shaded area the inter-quartile range.
    }
    \label{fig:error_vs_npert}
    
\end{figure}

%In the second numerical experiment we also 
We now turn our attention to detection of multiple perturbations. We fix $\lambda_1 = \SI{1530}{\nano\meter}$ and $\lambda_2 = \qty{1565}{\nano\meter}$, set the NF to \qty{6}{\decibel}, and perturb multiple sections simultaneously.
We run \num{1000} simulations, for each of which we generate random perturbations as follows:
First, we draw a perturbation count between \numlist{1;5} (inclusive) from a discrete uniform distribution.
Then, each perturbation is applied at a random section.
% We perform $1000$ independent cases, and evaluate each across $50$ independent fiber realizations while the amplifier has a noise figure of $\SI{6}{\decibel}$.
To label estimated locations as TPs, FPs and false negatives (FNs), we match them to the ground truths using linear sum assignment \cite{LinearSumAssig.} with a spatial tolerance threshold of $\SI{1}{\kilo\meter}$.
We report the localization error as a function of the number of perturbations.

\Cref{fig:error_vs_npert} (top) shows the TPs, TNs, FNs and localization error when applying up to five perturbations.
We achieve over \qty{90}{\percent} TPs for up to three perturbations. %, and never drop below \qty{85}{\percent}.
The noise in $R[l]$ increases as more perturbations occur.
As our peak detector threshold scales with the amplitude of $R[l]$, more peaks are discarded and the number of FNs grows.

\Cref{fig:error_vs_npert} (bottom) shows the median and IQR of the localization error, as the number of perturbations increases.
The median error grows from \qty{2.3}{\meter} for one perturbation to \qty{3.5}{\meter} for five.  
Similarly, the IQR increases from \qty{4.5}{\meter} to \qty{7.5}{\meter}.
% Although a single perturbation caused the highest proportion of false positives, the proportion of false positives and false negatives increase with an increased number of perturbations.
% (Bottom) The IQR broadens when multiple perturbations are introduced, while the median error remains approximately equivalent.
% The localization error remains sub-\qty{15}{\meter} for all multiple perturbation cases for a wavelength difference of $\qty{30}{\nano\meter}$.
We show one qualitative cross-correlation signal where five perturbations were present.
The figure shows the adaptive threshold as in \cref{fig:schematic:detector}, and a TP, FP or FN label for each peak.
As explained, each additional perturbation increases the noise in $R[l]$.
This noise reduces the accuracy of our parabolic fit to each peak, increasing the localization error.

% To explain the growing number of FNs, we decompose SOPAS as $a_m[l] = s_m[l] + n_m[l]$ where $s_m[l]$ contains the perturbation-induced peaks and $n_m[l]$ comes from EDFA noise.
% Writing $R[l]$ in terms of $s_m[l]$ and $n_m[l]$ reveals that more or larger perturbations locally amplify noise in $R[l]$.
% A higher noise floor increases $\hat\sigma$, leading to more FNs.

% Although not explicitly evident in our results, the number of FPs also increases when more perturbations are present. 
% Because $R[l]$ contains up to $M^2$ peaks, there is an increased probability that FPs happen to fall within the window bounded by the largest possible $t_1 - t_2$.

% \textcolor{red}{Discussion. For example, why do the error increase and TPs decrease with more perturbations? E.g. are the peaks closer together in this case, making them harder to detect? Or, is the threshold affected? What is happening here?}

\section{Conclusions}
We proposed a post-processing method that accurately localizes multiple fiber perturbations at the receiver using chromatic dispersion and the state of polarization.
Using only the C-band, simulation results demonstrate localization of up to \num{5} simultaneous perturbations with a median error of \qty{3.5}{\meter}.
% Our work expands the theoretical foundation for accurate fiber-optic multi-perturbation sensing systems.
% Our method can form a basis for distributed sensing in coherent receivers.
We believe our method can be used in coherent receivers to accurately pinpoint spatial perturbation profiles or multiple perturbations. 
% Our signals were exceptionally short (\SI{5}{\micro\second}).
% To  resulting in a very high event rate when multiple perturbations are applied.
% When two distinct events occur within a temporal window that is smaller than the maximum relative delay of the signals, additional false cross-correlation peaks will be formed.
% The dense temporal spacing of events in the simulation increases the probability of false positives.
% A peak amplitude depends on the $\tau_n$ of the fiber section and the SOP rotation of the perturbation.
% Together, they may potentially result in a weak peak, that combined with the noise floor, fails to clear the dynamic threshold.
% The probability of having a weak peak, and thus a false negative case, increases with the increased number of perturbations.
Future works include using a more detailed fiber model and an experimental validation of the proposed method, potentially considering different types of perturbations that replace the near-instantaneous random scramblers considered here.

% Our work has some limitations.
% First of all, it was evaluated on a linear single-span fiber model with no polarization-dependent loss, which is an ideal case.
% Additionally, the near-instantaneous, highly local perturbations we considered have no obvious equivalent in practice other than laboratory SOP scramblers.
% Most external events have some kind of spatiotemporal distribution instead.
% We assumed a noiseless transmitter, and receivers with perfect global phase compensation.
% Additionally, we assumed perfect synchronization between the receivers operating at different wavelengths.
% Finally, we counted FNs even for very minor perturbations, which are probably not of practical interest.
% While the current model assumes discrete perturbations and excludes practical constraints like nonlinear effects, 
% Future improvements could incorporate slow PMD drift and more realistic, time-varying physical perturbations. 
% Experimental validation using a more realistic fiber model and transceiver setup is essential to reveal practical challenges and confirm the practical feasibility of the sensing scheme. 
% Finally, the proposed method could be enhanced beyond detection and localization to measure the spatial distribution of perturbations.

% Future work could test the method on a more elaborate model or real fiber.
% The method could be expanded to infer the spatiotemporal distribution of real-life perturbations such as earthquakes.

% Results confirm five simultaneous events can be localized with sub-\qty{20}{\meter} error in a $\SI{50}{\kilo\meter}$ numerical fiber model.

\clearpage
\section{Acknowledgements}
This work was partially supported by the Swedish Research Council under grants no.~2025-04836 and 2025-03824 and by the project PINTO with file number NGF.1609.242.015 of the National Growth Fund (NGF) AiNed programme, which is financed by the Dutch Research Council (NWO). 

The authors would like to thank Prof. Gabriel Saavedra (Universidad de Concepción, Chile) and Prof. Tom Bradley (Eindhoven University of Technology, Eindhoven, the Netherlands) for fruitful discussions about the method presented in this work.

\defbibnote{myprenote}{%
}
\printbibliography[prenote=myprenote]

\vspace{-4mm}

%%%%%%%%%%%%%%%%%%%%%%%%%%%%%%%%%%%%%%%%%%%%%
%---------------------------------------------- End of Document -----------------------------------------------%
\end{document}